\documentclass[10pt,twocolumn]{article}
\usepackage{amsmath}
\usepackage{booktabs}
\usepackage{graphicx}
\usepackage{lineno}
\usepackage{url}
\usepackage{xcolor}

\usepackage[margin=0.75in]{geometry}
\usepackage{cite}
\usepackage{hyperref}
\graphicspath{{figures/out/}}

\newcommand{\figplaceholder}[1]{%
  \fbox{%
    \begin{minipage}[c][0.28\textheight][c]{0.90\linewidth}
      \centering
      Missing simulation figure\\[0.5em]
      \texttt{\detokenize{#1}}
    \end{minipage}%
  }%
}
\newcommand{\paperfigure}[2][]{%
  \IfFileExists{figures/out/#2}{\includegraphics[#1]{figures/out/#2}}{\figplaceholder{#2}}%
}

\begin{document}

\title{Dielectric-loaded guided-wave electro-optic modulator using bulk rubidium titanyl phosphate}
\author{David Trop\thanks{Department of Electrical and Electronics Engineering, Bar-Ilan University, Ramat Gan, Israel.} \and Boris Desiatov\thanks{Department of Electrical and Electronics Engineering, Bar-Ilan University, Ramat Gan, Israel. \texttt{boris.desiatov@biu.ac.il}}}
\date{}

\maketitle

\begin{abstract}
Rubidium titanyl phosphate (RTP) is widely used in bulk Pockels cells because it combines a useful
linear electro-optic response with low dielectric permittivity, high electrical resistivity,
negligible piezoelectric ringing, and high optical-damage tolerance. Its use in integrated or
guided-wave modulators has been limited by the absence of a mature thin-film RTP platform. Here we
propose a dielectric-loaded guided-wave modulator in which a silicon nitride strip on an unetched
bulk RTP crystal forms a near-surface optical mode that overlaps with side-electrode fields. The
analysis focuses on full-vector optical modes, GSG electrostatic and RF response, tensor-aware
electro-optic coupling, and device-level performance at 1064~nm.
\end{abstract}

\section{Introduction}

Electro-optic (EO) modulators are key components in optical communication, laser systems, and
ultrafast photonics, where they provide voltage-controlled phase or intensity modulation through
the Pockels effect. Lithium niobate (LN) and lithium tantalate (LT) dominate many guided-wave EO
platforms because they offer strong electro-optic coefficients and mature waveguide
processes~\cite{Wooten2000LithiumNiobateModulators,Boes2018LNOIReview,Wang2018IntegratedLNModulators,He2019HybridSiliconLNModulators,Powell2024ThinFilmLTModulators}.
However, their relatively large RF permittivity and piezoelectric response increase electrode
capacitance, microwave loss, and acoustic ringing, which can be limiting in short, high-voltage,
pulsed, or intracavity modulators. Rubidium titanyl phosphate (RTP, RbTiOPO$_4$) offers a
complementary material regime: bulk RTP Pockels cells are attractive for high-power and
high-repetition-rate systems because RTP combines low dielectric permittivity, high electrical
resistivity, negligible piezoelectric ringing, and high optical-damage tolerance. These properties
are especially valuable when the modulator must switch large optical powers cleanly, without
RF-driven acoustic artifacts~\cite{Albrecht2006RTP,Lebiush2000RTPQSwitch}.

The missing element is a practical guided-wave architecture. Previous RTP-family waveguides have
been demonstrated using ion exchange and reactive-ion-etched rib geometries, including
Cs$^+$-exchanged RTP channel waveguides and Mach--Zehnder structures
\cite{Cugat2013RTPChannelWaveguides,Butt2015RTPMachZehnderCsExchange,Choudhary2013RTPRibWaveguides}.
However, these approaches differ from the present dielectric-loaded architecture, which avoids ion
exchange and direct RTP etching while preserving access to the bulk crystal. RTP is not available
as a mature thin-film photonic platform comparable to thin-film LN or emerging thin-film LT
platforms
\cite{Boes2018LNOIReview,Wang2018IntegratedLNModulators,Powell2024ThinFilmLTModulators}, and
direct crystal etching or wafer transfer would add fabrication complexity and may compromise the
material advantages that motivate RTP in the first place. Consequently, RTP modulators have
remained mainly free-space devices, which limits mode control, footprint, and integration with
planar optical systems.

Here dielectric loading is used to create guided-wave modulation on an unpatterned bulk RTP
crystal. Related etchless loaded-waveguide strategies have shown that competitive EO performance
can be obtained without directly etching the active crystal
film~\cite{Qi2025LowIndexRibLoaded}; here, that fabrication philosophy is extended from thin-film
platforms to bulk anisotropic electro-optic crystals, where tensor-aligned interaction must be
co-designed with optical confinement.

We demonstrate that dielectric loading enables a bound near-surface optical mode in bulk RTP with
sufficient active-axis polarization and electrode overlap to achieve practical electro-optic
modulation efficiency. By co-designing crystal orientation, modal polarization, and electrode field
direction, the proposed structure enables guided-wave EO modulation in bulk RTP while preserving
its low-capacitance and high-power advantages without requiring thin-film processing or crystal
etching. The weak lateral confinement inherent to dielectric loading limits the achievable bend
radius, making the structure most suitable for straight or weakly curved phase modulators rather
than dense photonic integration.

In this work, we numerically evaluate a dielectric-loaded RTP guided-wave EO modulator in which a
silicon nitride (SiN) strip deposited on a polished bulk RTP surface locally increases the
effective index and produces a bound near-surface mode. Side electrodes apply a transverse field
across that mode, and the device is optimized by selecting crystal orientation,
guided-mode polarization, and electrode-field direction so that the optical mode samples a large
RTP tensor element while the electrical capacitance remains low. Unless otherwise stated, all
simulations are defined at $\lambda=1064~\mathrm{nm}$, a wavelength relevant to Nd:YAG and other
high-power solid-state laser systems.

\section{Device Concept and Orientation Co-Design}
\label{sec:concept}

\subsection{Dielectric-loaded geometry}

\begin{figure}[t]
\centering
\paperfigure[width=\linewidth]{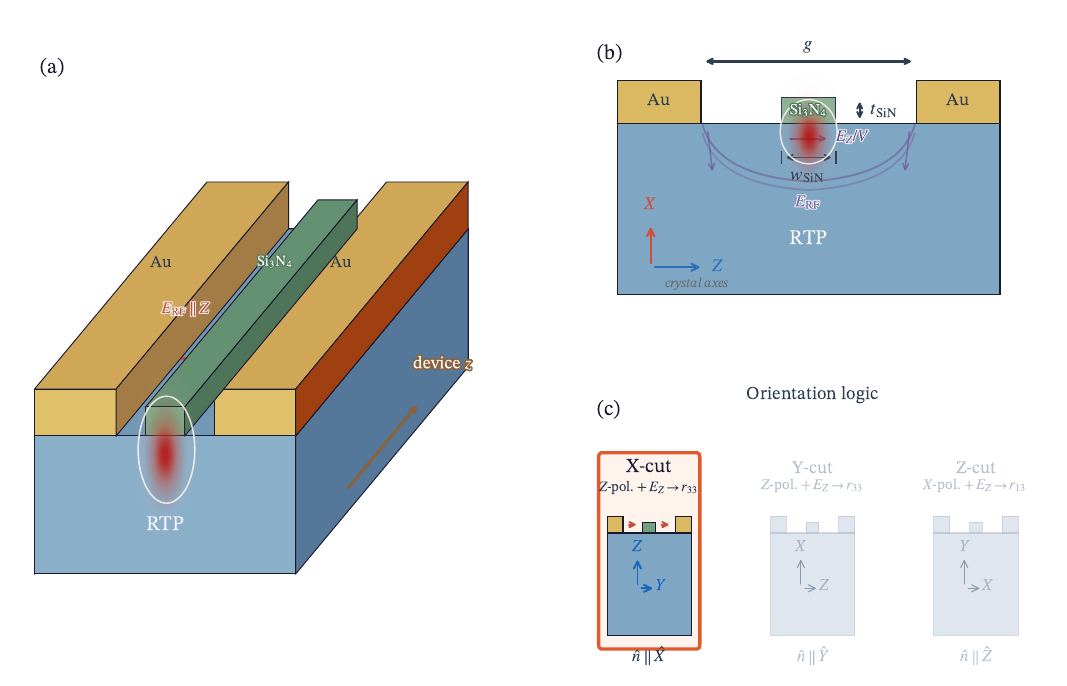}
\caption{Dielectric-loaded guided-wave RTP electro-optic modulator.
(a) Three-dimensional schematic of an unetched bulk RTP substrate, SiN loading strip, and side
electrodes, together with the near-surface optical mode and the device propagation direction.
(b) Cross-section defining the electrode gap $g$, SiN strip width $w_\mathrm{SiN}$, and SiN strip
thickness $t_\mathrm{SiN}$, together with the optical mode and the useful RF field component in
the RTP interaction region. (c) Orientation logic for the candidate crystal cuts. X-cut RTP is
used as the working layout because it provides a clean in-plane alignment between the guided-mode
polarization, the useful $E_Z$ field component, and the dominant $r_{33}$ interaction.}
\label{fig:device}
\end{figure}

Figure~\ref{fig:device} shows the proposed structure. A SiN strip of width $w_\mathrm{SiN}$ and
thickness $t_\mathrm{SiN}$ is deposited on an unpatterned RTP substrate. The upper cladding is air
for the primary design, with SiO$_2$ considered as a packaging-compatible secondary case. Gold side
electrodes are placed laterally around the strip with gap $g$ and optional lateral offset. For the
X-cut layout, optical propagation is defined along the in-plane device axis $z_\mathrm{prop}$,
taken parallel to the RTP crystal $Y$ axis, while the lateral electrode field is aligned primarily
with the RTP crystal $Z$ axis. Because the RTP itself is not etched or milled,
the propagation loss is not expected to be governed by RTP sidewall roughness. Instead, the
dominant fabrication-related optical-loss channels are the roughness of the patterned SiN loading
strip, the material quality of the deposited SiN film, and any scattering or absorption introduced
by the upper cladding. This makes the platform compatible with bulk RTP crystals while leveraging
established PECVD or LPCVD SiN processing~\cite{Beliaev2022SiNOpticalConstants}.

The SiN strip acts as a dielectric load rather than as a conventional fully embedded core. Since
$n_\mathrm{SiN}$ is larger than the principal RTP indices near 1064~nm, the strip raises the local
effective index of the near-surface region, following standard dielectric-waveguide guidance
principles~\cite{SnyderLove1983OpticalWaveguideTheory,Marcuse1991TheoryDielectricWaveguides}. In
this geometry, a bound mode satisfies
\begin{equation}
  n_\mathrm{RTP,bulk} < n_\mathrm{eff} < n_\mathrm{SiN},
  \label{eq:bound-mode-condition}
\end{equation}
where $n_\mathrm{RTP,bulk}$ is the relevant principal substrate index for radiation into the bulk.
If $n_\mathrm{eff}$ falls below the substrate index, the mode is leaky into the semi-infinite RTP
substrate.
For X-cut RTP, the relevant lower bound is not a single isotropic substrate index but the
appropriate principal RTP index, $n_X$, $n_Y$, or $n_Z$, depending on the crystal-axis content of
the mode and the radiation channel into the bulk.

Figure~\ref{fig:device}(a) defines the central design idea: the optical mode is formed by adding a
dielectric loading strip to a polished RTP surface, so the active crystal remains a bulk substrate
rather than a patterned thin film. Figure~\ref{fig:device}(b) shows the corresponding cross-section
used in the simulations. The optical mode is pulled toward the SiN-loaded surface, while the
side-electrode gap sets the useful field strength in the RTP interaction region. Thus the geometry
creates a controllable overlap between a micron-scale guided optical mode and a lateral RF field
without introducing RTP sidewalls. Figure~\ref{fig:device}(c) summarizes the orientation choice:
the device must simultaneously align the guided-mode polarization, the applied crystal-axis field,
and a large EO tensor element. This co-design requirement is the reason that crystal cut is treated
as part of the device geometry rather than as a fixed material label.

\subsection{Material parameters}

Table~\ref{tab:materials} lists the RTP material quantities used in the simulation model. The
principal refractive indices are calculated from the Sellmeier
equations of Mikami, Okamoto, and Kato~\cite{Mikami2009RTP}. With wavelength $\lambda$ expressed in
micrometers, the room-temperature dispersion is
\begin{subequations}
\label{eq:rtp-sellmeier}
\begin{align}
n_X^2(\lambda) &= 4.65575
  + \frac{0.04068}{\lambda^2 - 0.04750}
  + \frac{204.2586}{\lambda^2 - 130.7684},\\
n_Y^2(\lambda) &= 4.76892
  + \frac{0.04490}{\lambda^2 - 0.05130}
  + \frac{221.3309}{\lambda^2 - 134.2832},\\
n_Z^2(\lambda) &= 7.97109
  + \frac{0.06079}{\lambda^2 - 0.05968}
  + \frac{1234.6913}{\lambda^2 - 269.8094}.
\end{align}
\end{subequations}
The reported validity range is $0.5321 \leq \lambda \leq 3.1842~\mu\mathrm{m}$, which covers the
operating wavelength used here. At $\lambda=1.064~\mu\mathrm{m}$, Eq.~\eqref{eq:rtp-sellmeier}
gives $n_X=1.7657$, $n_Y=1.7743$, and $n_Z=1.8527$. The EO coefficients in
Table~\ref{tab:materials} initialize the tensor model. The SiN loading strip is modeled as an
isotropic dielectric with $n_\mathrm{SiN}=2.00$ at 1064~nm, representative of stoichiometric
silicon nitride films in the near infrared~\cite{Beliaev2022SiNOpticalConstants}. The RF
calculation uses the scalar permittivity approximation listed in the table.

\begin{table*}[t]
\caption{Material quantities used in the simulation model at
$\lambda=1064~\mathrm{nm}$.}
\label{tab:materials}
\centering
\small
\resizebox{\textwidth}{!}{%
\begin{tabular}{@{}p{0.34\linewidth}p{0.56\linewidth}@{}}
\toprule
Quantity & Value or model \\
\midrule
Optical indices at $1.064~\mu$m & $n_X=1.7657$, $n_Y=1.7743$, $n_Z=1.8527$~\cite{Mikami2009RTP,Nikogosyan2005NonlinearOpticalCrystals} \\
SiN loading-strip index & Isotropic $n_\mathrm{SiN}=2.00$ at 1064~nm, representative of stoichiometric SiN~\cite{Beliaev2022SiNOpticalConstants} \\
Dominant EO coefficient & $r_{33}=33.0\pm3.3$~pm/V~\cite{Nikogosyan2005NonlinearOpticalCrystals}; device reports give $\sim35$~pm/V~\cite{Albrecht2006RTP} \\
Secondary EO coefficients & $r_{13}=10.9\pm1.1$~pm/V, $r_{23}=15.0\pm1.5$~pm/V~\cite{Nikogosyan2005NonlinearOpticalCrystals} \\
RF permittivity & Scalar $\epsilon_r\approx13$ used for the quasi-static RF estimates~\cite{Fragemann2001RTPModulator}; related low-frequency RTP dielectric data are reported in Ref.~\cite{Perumal2020RTPDielectric} \\
RF loss and high-power metrics & Reported through RF attenuation and optical-field proxy metrics in the device simulations \\
\bottomrule
\end{tabular}
}
\end{table*}

\subsection{Crystal orientation}
\label{sec:orientation}

Efficient side-electrode modulation requires the useful RF field component, the optical
polarization, and the EO tensor to be aligned. RTP is biaxial, so a scalar polarization shortcut is
not sufficient; the simulation must track fields in the crystal-axis basis.

The preferred starting point is X-cut RTP, with the surface normal parallel to the crystal $X$
axis. In this geometry the $Y$ and $Z$ axes lie in the chip plane, so a lateral electrode pair can
produce a dominant RF field component along $Z$. If the guided mode is also predominantly
$Z$-polarized in the RTP region, the device accesses the large $r_{33}$ coefficient.

To verify that this choice is not only a geometric assumption, we compare X-cut, Y-cut, and Z-cut
RTP using the same dielectric-loaded waveguide cross-section. The orientation comparison, shown in
Fig.~\ref{fig:orientation_modes}, evaluates the guided mode, the crystal-axis field fractions, and
a tensor-aligned orientation metric for each cut. This comparison identifies whether the selected
cut simultaneously supports a bound near-surface mode and provides access to the desired EO tensor
projection.

The relevant design metric is therefore not the peak optical field, not $|E_{\mathrm{RF}}|$, and
not the material coefficient alone, but the tensor-weighted overlap between the guided optical mode
and the active RF field component inside RTP.

In this orientation screen, X-cut denotes an RTP surface normal along the crystal $X$ axis. This
places the $Y$ and $Z$ axes in the chip plane, so side electrodes can drive a useful $E_Z$ field
component and a predominantly $Z$-polarized guided mode can access the intended $r_{33}$
interaction. Y-cut RTP, with surface normal along $Y$, can also access a $Z$-polarized,
$E_Z$-driven $r_{33}$ interaction in principle, but it is not the working layout used for the
device design. Z-cut RTP, with surface normal along $Z$, maps the chosen in-plane side-electrode
drive onto weaker effective tensor paths and, for the same SiN-loaded cross-section, rotates the
substrate index axes so that the horizontally polarized mode is no longer vertically confined near
the surface. The subsequent optical, RF, and EO simulations therefore focus on X-cut RTP.

\section{Numerical Method}
\label{sec:method}

\subsection{Simulation framework and optical modes}

All numerical simulations were performed using Tidy3D by Flexcompute~\cite{FlexcomputeTidy3D}.
The same simulation framework is used for the optical eigenmode calculations, electrostatic and RF
field extraction, and tensor-aware electro-optic perturbation analysis. All simulations use a
shared cross-section containing the anisotropic RTP substrate, the SiN loading strip, the upper
cladding, and the Au side electrodes. The same geometry is used for orientation screening, optical
modes, electrostatics, EO perturbation, and tolerance sweeps. For each design point, the
unperturbed optical mode is solved first, the electrode field is then computed and normalized to the
applied voltage, and the resulting crystal-axis Pockels perturbation is mapped back into the
optical solver to extract $\partial n_\mathrm{eff}/\partial V$. Capacitance is obtained from
electrostatic charge or energy integration using the same GSG electrode geometry.

The optical mode is computed with a full-vector eigenmode solver using the anisotropic RTP
permittivity tensor in the crystal-axis basis. For each geometry, we extract
$n_\mathrm{eff}$, group index, mode area, number of bound modes, confinement fractions in
RTP/SiN/cladding, mode-centroid depth, and the $X$-, $Y$-, and $Z$-polarized field-energy fractions
inside RTP. Mesh and boundary padding are converged until $n_\mathrm{eff}$ and the active-axis
polarization fraction are stable. In practice, the optical mesh and simulation window were accepted
when further refinement changed $n_\mathrm{eff}$ by less than $1\times10^{-4}$, changed the
RTP/SiN confinement fractions by less than $1$ percentage point, and changed the active-axis
polarization fraction by less than $0.5$ percentage point. The electrostatic/RF mesh was accepted
when the extracted capacitance and characteristic impedance changed by less than $2$\% under mesh
refinement or domain padding. These tolerances are substantially smaller than the geometry-driven
changes discussed below and correspond to less than about $1$--$2$\% uncertainty in the final
$V_\pi L$ values.

\subsection{Electrostatic, RF, and EO extraction}

The electrode problem is treated first in the electrostatic limit for short lumped-device figures of
merit. Fields are normalized to $V_\mathrm{applied}=1~\mathrm{V}$, and capacitance per unit length
$C'$ is extracted from charge or energy integration. A quasi-TEM RF calculation is used to extract
impedance, RF effective index, and conductor/dielectric attenuation for a traveling-wave extension,
following the standard coplanar-waveguide treatment of surface transmission lines
\cite{Wen1969CoplanarWaveguide,Simons2001CoplanarWaveguide}.

In the principal-axis basis, the linear EO perturbation is
\begin{equation}
  \Delta\left(\frac{1}{n_i^2}\right)=\sum_j r_{ij}E_j ,
  \label{eq:pockels}
\end{equation}
where $i$ indexes the optical principal axis and $j$ indexes the applied field component. For the
X-cut design, the dominant term is expected to be
\begin{equation}
  \Delta n_Z(\mathbf{r}) \approx -\frac{1}{2}n_Z^3 r_{33}E_Z(\mathbf{r}).
  \label{eq:dn}
\end{equation}
The spatially varying perturbation is mapped into the optical solver to extract the voltage
response of the guided mode,
$\partial n_\mathrm{eff}/\partial V$, from which
\begin{equation}
  V_\pi L = \frac{\lambda}{2\left|\partial n_\mathrm{eff}/\partial V\right|}.
  \label{eq:vpi_from_dneff}
\end{equation}
is obtained for a single-arm phase shifter. For rapid sweeps, the mode-weighted EO overlap
efficiency is defined as
\begin{equation}
  \eta_\mathrm{EO}
  =
  \frac{\displaystyle\int_{\mathrm{RTP}} W_\mathrm{opt}(\mathbf{r})E_Z(\mathbf{r})\,dA}
       {\displaystyle V/g},
  \label{eq:eta_eo}
\end{equation}
where $W_\mathrm{opt}$ is normalized over the RTP region. This overlap is used to interpret and
screen the geometry space, while the final results are taken from the direct
$\partial n_\mathrm{eff}/\partial V$ calculation.
As a numerical sanity check, the EO perturbation was evaluated in both positive and negative
voltage directions and retained only when the fitted $n_\mathrm{eff}(V)$ slope was linear to within
2\% over the applied voltage range. A uniform-field limit was also checked against the analytic
Pockels relation $\Delta n_i=-(1/2)n_i^3 r_{ij}E_j$; agreement within a few percent was taken as
the acceptance criterion for the tensor mapping and sign convention. These validation checks are
used to assign the reported $V_\pi L$ uncertainty to numerical convergence rather than to the much
larger geometry tradeoffs shown in Figs.~\ref{fig:eo_performance} and~\ref{fig:design_tradeoffs}.

\section{Optical Guided Modes}
\label{sec:optical}

\subsection{Guided-mode window and mode area}

\begin{figure}[t]
\centering
\paperfigure[width=\linewidth]{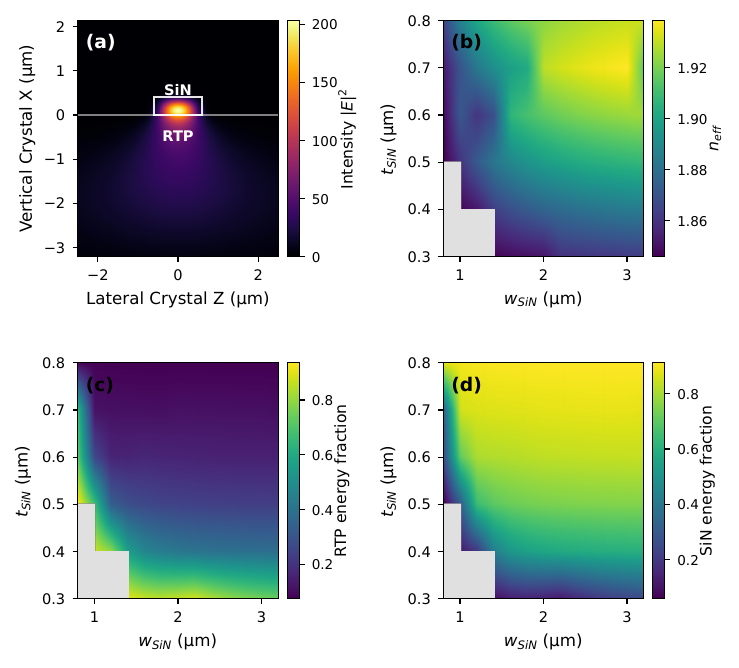}
\caption{Optical guided mode of the dielectric-loaded RTP waveguide.
(a) Fundamental mode intensity at the selected geometry. (b) Effective-index map versus SiN width
and thickness. (c) RTP energy fraction over the same geometry space. (d) SiN energy fraction over
the same geometry space. Light-gray regions indicate geometries for which no bound mode is
identified.}
\label{fig:optical_mode}
\end{figure}

The SiN-loaded RTP surface supports a bound mode when $n_\mathrm{eff}$ remains above the relevant
RTP substrate index and below the SiN index. Figure~\ref{fig:optical_mode} shows the fundamental
mode profile at the selected geometry and the dependence of $n_\mathrm{eff}$ on $w_\mathrm{SiN}$
and $t_\mathrm{SiN}$, thereby identifying the bound-mode region and separating it from leaky or
weakly confined regimes.

Figure~\ref{fig:optical_mode}(a) confirms that the dielectric load creates a near-surface guided
mode rather than a deeply buried substrate mode. The field maximum remains close to the SiN/RTP
interface, but a substantial part of the mode samples the RTP underneath the strip, which is the
region where the Pockels interaction is later driven. Figure~\ref{fig:optical_mode}(b) shows how
the effective index rises as the SiN strip becomes wider or thicker. The light-gray regions mark
geometries where the dielectric load is too weak to support the target bound mode, while the
high-index side of the map indicates stronger confinement and a larger fraction of the optical
energy in the loading strip.

The confinement maps in Fig.~\ref{fig:optical_mode}(c,d) make the central optical compromise more
explicit. Increasing the SiN loading generally improves lateral confinement, but it also transfers
more energy from RTP into SiN. The selected geometry is therefore not chosen at the strongest
possible loading; it is placed in a window where the mode is bound and compact while still
retaining meaningful optical energy in the RTP substrate. Beyond mode existence, this window also
sets the onset of higher-order modes and the effective mode area relevant to high-power operation.

\subsection{Orientation-dependent modal alignment}

\begin{figure*}[t]
\centering
\paperfigure[width=\textwidth]{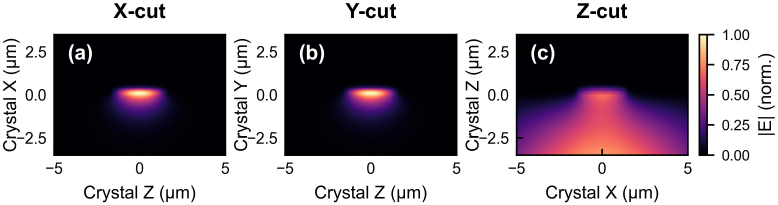}
\caption{Orientation-dependent guided modes and optical alignment in dielectric-loaded RTP.
(a--c) Fundamental optical modes for X-cut, Y-cut, and Z-cut RTP simulated using the same
SiN-loaded cross-section. X-cut is retained as the working geometry for the selected surface and
side-electrode layout because it provides clean alignment between optical confinement, modal
polarization, and the useful $E_Z$ field component. Y-cut provides an additional control that can
also access an $E_Z$-driven $r_{33}$ path in principle, while Z-cut maps the chosen drive onto
weaker effective tensor paths and does not preserve the same near-surface confinement.}
\label{fig:orientation_modes}
\end{figure*}

The orientation study compares X-cut, Y-cut, and Z-cut RTP using the same SiN-loaded cross-section.
Figure~\ref{fig:orientation_modes} shows the fundamental guided mode for each cut. This comparison
separates two requirements that must be satisfied simultaneously: the structure must support a
bound near-surface optical mode, and the modal polarization must remain compatible with a
crystal-axis field component that can later be driven efficiently by the side electrodes.

For the X-cut geometry, the crystal $Z$ axis lies in the device plane. The side-electrode field
can therefore drive a useful $E_Z$ component, while the guided mode retains a large
$Z$-polarized field fraction inside RTP. This combination enables the intended $r_{33}$
interaction. The Y-cut and Z-cut cases provide useful controls: they may also support guided modes,
but their crystal-axis mapping changes the balance between optical confinement, modal polarization,
and side-electrode field direction. Y-cut can also access an $E_Z$-driven $r_{33}$ interaction in
principle, whereas Z-cut changes both the optical cutoff condition and the tensor projection. In a
biaxial crystal, changing the cut rotates the substrate indices $(n_X,n_Y,n_Z)$ relative to the
surface plane and to the dominant horizontal modal field. For the Z-cut case, this index
orientation weakens the effective contrast needed for the horizontally polarized near-surface mode:
the field spreads laterally and penetrates deeply into the RTP substrate instead of remaining
localized under the SiN load. Z-cut is therefore rejected not only because the chosen
side-electrode drive maps onto weaker effective tensor paths, but also because the vertical optical
confinement collapses for this cross-section. X-cut is retained as the working design for the
subsequent EO and RF analysis because it gives the cleanest combined optical and tensor alignment
in the selected surface-electrode layout.

After the orientation is fixed, local sweeps around the X-cut design point are used to confirm
that the active-axis fraction remains stable across the relevant SiN width and thickness range.

Figure~\ref{fig:orientation_modes}(a) shows the desired X-cut case, where the guided field remains
well localized near the dielectric load and the dominant optical polarization can be paired with an
in-plane $E_Z$ drive. Figure~\ref{fig:orientation_modes}(b) shows that Y-cut RTP can also support a
guided mode in the same physical cross-section, but the crystal-axis mapping changes how the modal
polarization and side-electrode field project onto the EO tensor. Figure~\ref{fig:orientation_modes}(c)
provides the complementary Z-cut control. Here the problem is more severe than tensor alignment
alone: rotating the crystal axes places the relevant substrate indices in a different relation to
the surface and to the horizontally polarized field, so the same SiN load no longer produces a
compact near-surface mode. The optical field becomes laterally diffuse and leaks deeply into the
substrate, indicating a collapse of vertical confinement for this cross-section. The orientation
comparison therefore rules out both a scalar-index optical selection criterion and a
coefficient-only EO selection criterion. A useful RTP guided-wave modulator must be selected by the
combined optical mode, electrode field direction, and EO tensor projection, which motivates the
X-cut choice used below.

\section{Electrode Design and Electrostatic Field}
\label{sec:electrodes}

\subsection{Side-electrode geometry}

The baseline electrical configuration uses a coplanar ground-signal-ground (GSG) electrode
structure on the same surface as the SiN strip. The signal electrode is centered over the optical
interaction region, while the two ground electrodes are placed laterally on either side. The
principal swept variables are the electrode gap $g$, the signal width $w_{\mathrm{sig}}$, the
ground width, the metal thickness, and the lateral offset relative to the optical mode. Because the
useful electro-optic interaction is tensor-selective, the electrical analysis is expressed in the
crystal-axis basis and focuses on the normalized useful field component, $E_Z/V$, rather than on
total field magnitude alone.

\subsection{Useful RF field distribution}

The electrostatic design is governed by the overlap between the useful RF field component and the
guided optical mode inside RTP. The normalized useful field distribution is therefore interpreted
together with optical-mode contours, which show whether the field is concentrated at the mode
centroid or instead peaks near metal corners and outside the main interaction region.

\begin{figure}[t]
\centering
\paperfigure[width=\linewidth]{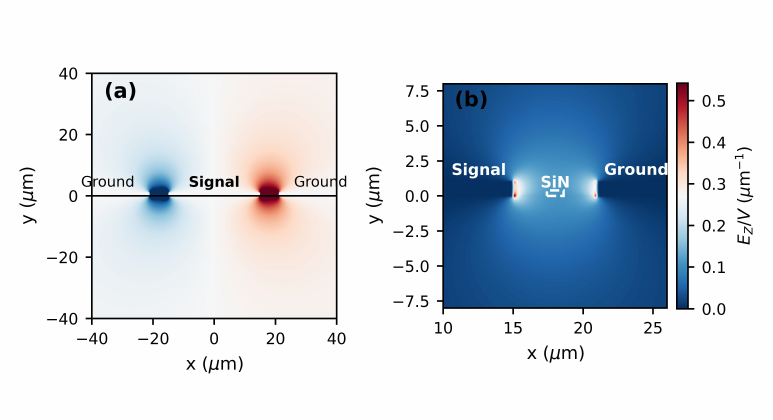}
\caption{Electrostatic field distribution in the GSG electrode structure.
(a) Useful RF field component $E_Z/V$ in the full cross-section.
(b) Zoomed view of the signal-ground interaction region showing the field concentration around the
SiN-loaded RTP surface.}
\label{fig:rf_field}
\end{figure}

Figure~\ref{fig:rf_field}(a) shows the full electrostatic cross-section and separates the useful
field component from the total fringing field of the GSG electrodes. The highest fields occur near
the metal edges, as expected for a coplanar geometry, but the relevant question for modulation is
whether the $E_Z/V$ component reaches the optical mode in RTP. Figure~\ref{fig:rf_field}(b)
answers this locally: in the signal-ground interaction region, the useful field penetrates the
SiN-loaded RTP surface and overlaps the guided mode rather than remaining confined to the metal
corners. This field distribution justifies the side-electrode geometry and sets up the later gap
sweep: reducing $g$ should increase EO efficiency, but it also brings lossy metal closer to the
guided optical field.

\subsection{Capacitance and RF modal properties}

Low capacitance is one of the main motivations for guided-wave RTP modulators, but electrostatic
overlap alone does not determine the usable modulation bandwidth. The RF analysis therefore
extracts the capacitance per unit length, the microwave effective index, and the characteristic
impedance of the electrode-loaded structure as functions of the main electrode geometry. Comparing
the microwave effective index with the optical group index indicates whether the proposed phase
shifter remains close to velocity matched over practical interaction lengths.

To characterize the GSG electrode structure beyond a single design point, a two-dimensional RF
sweep is performed over signal width $w_{\mathrm{sig}}$ and electrode gap $g$. This sweep maps the
characteristic impedance across the practical design space and identifies a continuous set of
geometries near $50~\Omega$. A selected operating point on this matching trajectory is then
characterized versus frequency to extract the microwave effective index, RF loss, and impedance
stability. This separates the static geometry-selection problem from the frequency-dependent RF
behavior of the final line.

Figure~\ref{fig:rf_properties} summarizes the main microwave properties of the structure. The
impedance map identifies a practical matched geometry, while the frequency sweeps show whether that
geometry maintains suitable microwave index and RF loading over the band of interest. These results
are used together with the electro-optic analysis to estimate the practical interaction length and
bandwidth of the proposed device.

\section{Electro-Optic Performance}
\label{sec:eo}

\subsection{Overlap and $V_\pi L$}

\begin{figure}[t]
\centering
\paperfigure[width=\linewidth]{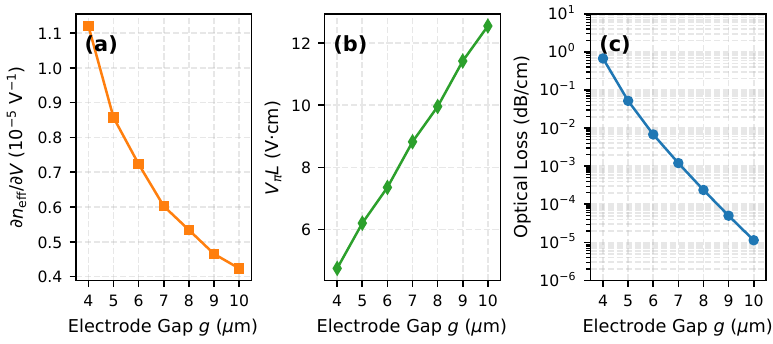}
\caption{Electro-optic response versus electrode gap for the dielectric-loaded RTP modulator at
fixed $w_\mathrm{SiN}=1.2~\mu\mathrm{m}$, $t_\mathrm{SiN}=400~\mathrm{nm}$, and
$w_\mathrm{sig}=7.75~\mu\mathrm{m}$.
(a) Voltage derivative of the effective index, $|\partial n_\mathrm{eff}/\partial V|$, obtained
from the tensor-aware perturbation of the anisotropic RTP permittivity. (b) Corresponding
half-wave-voltage-length product, $V_\pi L$, calculated using Eq.~\eqref{eq:vpi_from_dneff}.
(c) Electrode-induced optical loss versus gap, defined as the excess propagation loss caused by the
metal electrodes relative to the no-metal waveguide. Together, the panels show the expected
trade-off: reducing the electrode gap improves modulation efficiency but increases the optical-loss
penalty from the electrodes.}
\label{fig:eo_performance}
\end{figure}

\begin{figure}[t]
\centering
\paperfigure[width=\linewidth]{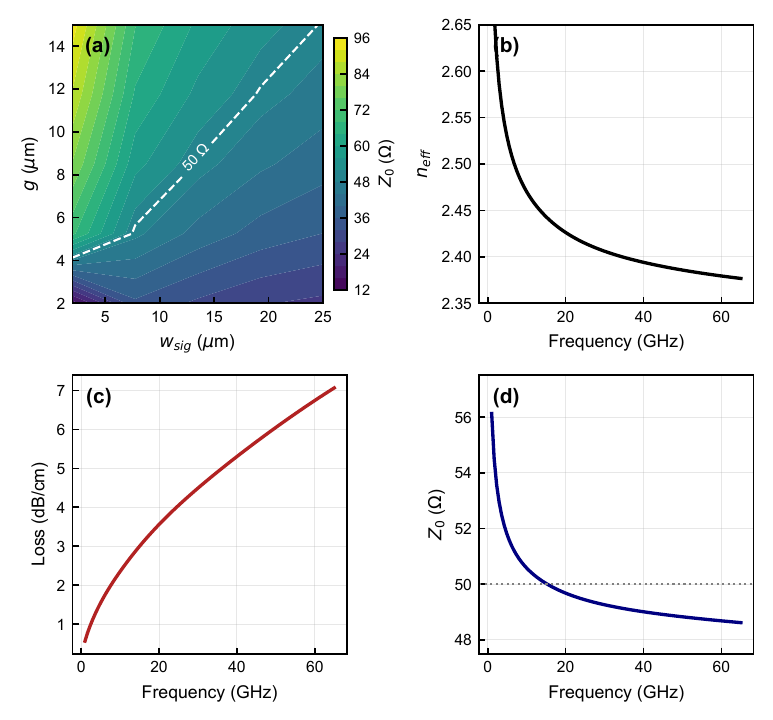}
\caption{RF-wave properties of the GSG electrode-loaded RTP modulator.
(a) Characteristic impedance in the $(w_{\mathrm{sig}}, g)$ design space, with the dashed contour
marking geometries near $50~\Omega$. (b) Microwave effective index versus frequency for the
selected operating point. (c) RF loss versus frequency. (d) Characteristic impedance versus
frequency.}
\label{fig:rf_properties}
\end{figure}

The electro-optic response of the proposed structure is governed by a tensor-weighted interaction
between the guided optical mode and the useful RF field inside RTP.
The modulation efficiency is set by how strongly the optical field samples the crystal-axis
component driven by the electrode field.
Accordingly, final $V_\pi L$ values are extracted from the tensor-aware perturbation of the
anisotropic RTP permittivity rather than from a scalar overlap model.
To connect the RF-field distribution to the final electro-optic response, we first evaluate the
mode-weighted overlap efficiency defined in Eq.~\eqref{eq:eta_eo}. This dimensionless quantity
measures how efficiently the applied voltage is converted into the tensor-relevant RF field
sampled by the guided mode and is used to interpret the gap dependence.
The quantity plotted in Fig.~\ref{fig:eo_performance}, $\partial n_\mathrm{eff}/\partial V$, is the
slope of the voltage-dependent
effective index of the guided optical mode. It is the direct output of the tensor perturbation
calculation and determines the phase shift according to
$\Delta\phi=(2\pi/\lambda)(\partial n_\mathrm{eff}/\partial V)VL$. Thus Fig.~\ref{fig:eo_performance}
is calculated to connect the tensor-aware index response directly to the final phase-shifter
metric. As the electrode gap increases, the useful RF field sampled by the optical mode decreases,
which reduces $|\partial n_\mathrm{eff}/\partial V|$ and increases $V_\pi L$.
The same gap sweep also tracks electrode-induced optical loss, defined as the excess optical
propagation loss introduced by the nearby metal electrodes relative to the corresponding
SiN-loaded RTP waveguide without lossy metal. This optical absorption penalty is distinct from the
RF conductor loss reported in Fig.~\ref{fig:rf_properties}.
Figure~\ref{fig:eo_performance} isolates the electrode-gap dependence using the fixed optical
geometry $w_\mathrm{SiN}=1.2~\mu\mathrm{m}$ and
$t_\mathrm{SiN}=400~\mathrm{nm}$, with $w_\mathrm{sig}=7.75~\mu\mathrm{m}$.
The selected point in this one-parameter sweep, $g=5.25~\mu\mathrm{m}$, is the same validated
geometry used for the performance summary in Table~\ref{tab:performance} and for the operating
window in Fig.~\ref{fig:design_tradeoffs}.

Figure~\ref{fig:eo_performance}(a) shows the direct voltage response of the guided mode. As the
electrode gap is reduced, the optical mode samples a larger useful RF field and
$|\partial n_\mathrm{eff}/\partial V|$ increases. Figure~\ref{fig:eo_performance}(b) converts the
same index response into the single-arm $V_\pi L$ metric, making the expected inverse relation
between RF-field strength and drive voltage explicit. The smallest gap therefore gives the best
pure EO efficiency. Figure~\ref{fig:eo_performance}(c), however, shows why that point is not
automatically the preferred design: moving the metal closer to the waveguide increases
electrode-induced optical loss. The gap sweep therefore establishes a genuine device tradeoff,
not simply a voltage-minimization problem.

In Fig.~\ref{fig:rf_properties}(a), the $50~\Omega$ contour identifies a family of GSG geometries
rather than a single isolated solution. This is important because the optical design and the EO
gap dependence impose additional constraints: the final electrode gap must be compatible with both
modulation efficiency and electrode-induced optical loss. Figure~\ref{fig:rf_properties}(b) shows
that the selected matched point has a microwave effective index in the same range as the guided
optical group index, so the line is not grossly velocity mismatched for millimeter-scale phase
shifters. Figures~\ref{fig:rf_properties}(c,d) then check the frequency dependence of that same
operating point. The RF loss increases with frequency, while the impedance remains near the
designed value, indicating that the selected geometry is a plausible RF load rather than only a
static capacitance minimum. Together, these panels support using the matched GSG point as the
electrical reference for the EO and device-level comparisons.

Accordingly, the electro-optic analysis combines the optical and electrical solutions through the
tensor-weighted overlap, the voltage derivative of the effective index,
$\partial n_\mathrm{eff}/\partial V$, and the resulting half-wave-voltage-length product,
$V_\pi L$. The tensor-weighted overlap is used to screen the geometry space, whereas the final
$V_\pi L$ values are extracted from the direct electro-optic perturbation calculation described in
Section~\ref{sec:method}. This separation keeps the design sweeps efficient
while preserving a rigorous final performance extraction.

\subsection{Design tradeoffs and operating-point selection}

The final device assessment combines electro-optic efficiency with the optical-confinement and
high-power proxy metrics of the dielectric-loaded structure. The preferred operating point is not
the minimum-$V_\pi L$ geometry alone. Instead, it is selected by balancing the voltage-length
product against confinement in the bulk RTP region and optical exposure near material interfaces.
We define $\Gamma_\mathrm{interface}$ as the fraction of optical energy located within 50~nm of
the RTP surface interface sampled by the optical-mode calculation. This quantity is used as a
qualitative high-power compatibility proxy: lower values indicate reduced optical concentration
near the loaded-crystal surface, but it is not a substitute for a measured damage threshold.
Although the minimum-$V_\pi L$ geometry occurs at the smallest electrode gap, the final design
intentionally operates at a slightly larger gap in order to reduce electrode-induced optical loss
while retaining meaningful RTP confinement and acceptable RF characteristics.

\begin{figure}[t]
\centering
\paperfigure[width=\linewidth]{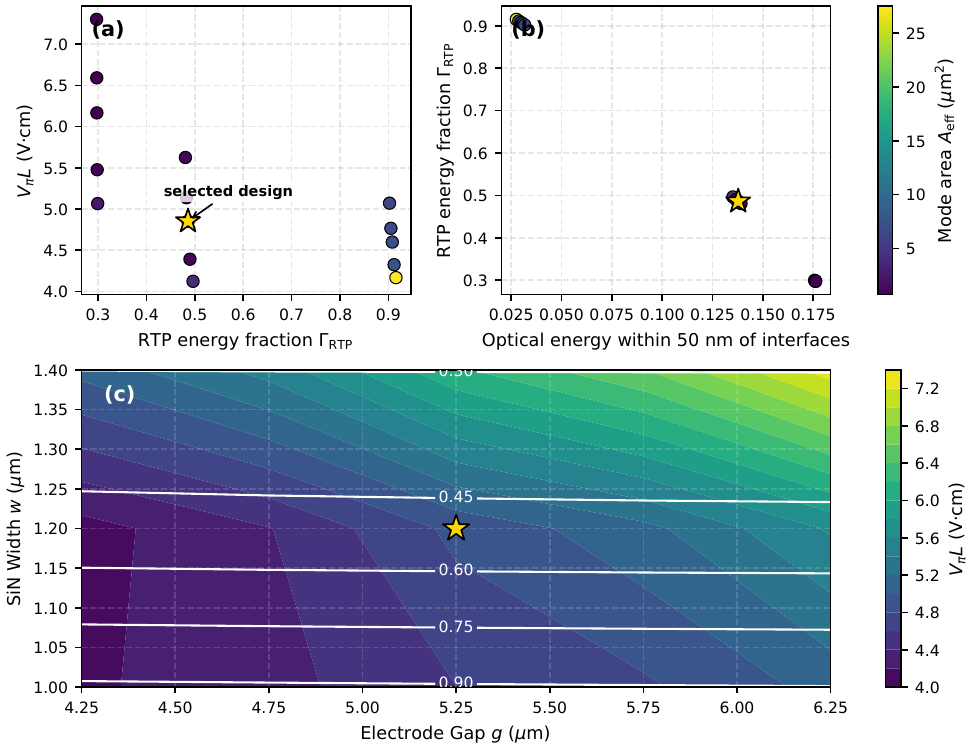}
\caption{Design tradeoffs and operating-point selection for the dielectric-loaded RTP modulator.
(a) Half-wave-voltage-length product as a function of RTP confinement, with marker color
indicating optical mode area. (b) RTP confinement versus optical energy within 50~nm of material
interfaces, illustrating the tradeoff between RTP guidance and surface-interface exposure.
(c) Design-space map in the $(g,w_\mathrm{SiN})$ plane, showing calculated $V_\pi L$ by color and
RTP confinement as contours. The selected operating point is marked by the star.}
\label{fig:design_tradeoffs}
\end{figure}

Figure~\ref{fig:design_tradeoffs}(a) places the EO metric directly against RTP optical
confinement. This view is more informative than a voltage-only sweep because it shows whether a
lower $V_\pi L$ is obtained by sacrificing the bulk-crystal character of the mode. The selected
point lies in the balanced part of this space: it retains meaningful RTP energy while keeping the
single-arm voltage-length product practical. Figure~\ref{fig:design_tradeoffs}(b) separates a
second design axis, the exposure of the optical mode to nearby material interfaces. Designs with
similar RTP confinement can still differ in how much energy is concentrated within 50~nm of the
loaded RTP surface, which is why this panel is used as a qualitative high-power
compatibility check. Figure~\ref{fig:design_tradeoffs}(c) returns these tradeoffs to the physical
geometry plane. The selected star is not located at the absolute minimum of the color map; instead,
it sits in an operating window where $V_\pi L$, RTP confinement, RF matching, and
electrode-induced loss can all be satisfied simultaneously.

\subsection{Optimized design point}

The optimized design point is selected by balancing electro-optic efficiency, capacitance, optical
confinement, active-axis polarization purity, metal separation, and fabrication tolerance. It is
therefore defined not by the minimum $V_\pi L$ alone, but by the best overall trade-off among
tensor-selective modulation efficiency, electrical loading, and fabrication robustness.
The selected geometry used to connect Figs.~\ref{fig:optical_mode}--\ref{fig:design_tradeoffs}
is X-cut RTP with $w_\mathrm{SiN}=1.2~\mu\mathrm{m}$, $t_\mathrm{SiN}=400~\mathrm{nm}$,
electrode gap $g=5.25~\mu\mathrm{m}$, and GSG signal width
$w_\mathrm{sig}=7.75~\mu\mathrm{m}$. This electrode point lies on the near-$50~\Omega$ RF
matching trajectory, with $Z_0\approx49~\Omega$ and microwave effective index
$n_\mathrm{RF}\approx2.42$--2.43.

Table~\ref{tab:performance} summarizes the key quantities required to describe the optimized RTP
design at 1064~nm. The simulated propagation-loss entry reports the electrode-induced optical loss
at the selected gap, relative to the corresponding no-metal dielectric-loaded RTP waveguide in the
lossless-material model. Because the present simulations do not include a measured deposited-SiN
absorption and roughness model, we also quote an engineering total-loss budget by adding a
representative $1~\mathrm{dB/cm}$ allowance for SiN film absorption and scattering to the simulated
electrode-induced loss. This gives an estimated total propagation loss of
$\sim1.3~\mathrm{dB/cm}$, corresponding to $\sim86$\% transmission through a 5~mm phase shifter and
$\sim74$\% through a 10~mm phase shifter, before fiber/chip coupling losses.

\begin{table*}[t]
\caption{Performance summary for the optimized RTP design at
$\lambda=1064~\mathrm{nm}$.}
\label{tab:performance}
\centering
\resizebox{\textwidth}{!}{%
\begin{tabular}{lll}
\toprule
Quantity & Value & Unit \\
\midrule
SiN width $w_\mathrm{SiN}$ & 1.2 & $\mu$m \\
SiN thickness $t_\mathrm{SiN}$ & 400 & nm \\
Electrode gap $g$ & 5.25 & $\mu$m \\
GSG signal width $w_\mathrm{sig}$ & 7.75 & $\mu$m \\
Characteristic impedance $Z_0$ & $\approx49.0$ & $\Omega$ \\
Microwave effective index $n_\mathrm{RF}$ & $\approx2.43$ & -- \\
Effective index $n_\mathrm{eff}$ & $\approx1.858$ & -- \\
Optical mode area $A_\mathrm{eff}$ & $\approx1.11$ & $\mu$m$^2$ \\
Optical energy in RTP & $\approx48.6$ & \% \\
Optical energy in SiN & $\approx49.3$ & \% \\
Active-axis field fraction in RTP & $\approx99.9$ & \% \\
$|\partial n_\mathrm{eff}/\partial V|$ & $\approx1.10\times10^{-5}$ & V$^{-1}$ \\
Single-arm $V_\pi L$ & $\approx4.85$ & V\,cm \\
Electrode-induced propagation loss & $\approx0.286$ & dB/cm \\
Estimated total propagation-loss budget & $\approx1.3$ & dB/cm \\
Capacitance $C'$ & $\approx167$ & fF/mm \\
Estimated RC bandwidth (1~mm, 50~$\Omega$) & $\approx19$ & GHz \\
\bottomrule
\end{tabular}
}
\end{table*}

\section{Comparison with Bulk RTP and Device-Level Performance}
\label{sec:benchmark}

Representative bulk RTP Pockels-cell operation provides the practical reference for voltage,
aperture, crystal length, and capacitance. This comparison tests whether micron-scale guided-wave
confinement can reduce the drive-voltage burden while preserving the bulk-compatible fabrication
and high-power motivation of RTP.

\subsection{Comparison with bulk RTP Pockels cells}

Conventional RTP Pockels cells use millimeter-scale optical apertures and electrode spacings, which
enable high optical-power handling but typically require high half-wave voltages. In the proposed
guided-wave geometry, the optical mode is confined near the surface and the electrode spacing can be
reduced to the micrometer scale, producing a much larger electric field per applied volt. The
comparison focuses on voltage reduction for phase modulation in a straight interaction region rather
than on dense integration. Table~\ref{tab:bulk_comparison} summarizes the comparison with
representative bulk RTP Pockels-cell operation.

As a practical bulk-device reference, commercial RTP EO cells are typically millimeter-aperture,
kilovolt-drive components rather than tightly confined guided-wave devices. For example, Raicol
reports a half-wave voltage of 3.6~kV for an RTP EO cell of size $9 \times 9 \times 10$~mm,
clear apertures from $1.5 \times 1.5$~mm$^2$ to $15 \times 15$~mm$^2$, crystal lengths up to
50~mm, and transmission above 98.5\% at 1064~nm
\cite{RaicolRtpEoCellWeb,RaicolRtpEoCellDatasheet}. These values are consistent with the role of
bulk RTP as a high-power, large-aperture electro-optic platform, as also reflected in RTP
Q-switch and EO-cell studies~\cite{Lebiush2000RTPQSwitch,Albrecht2006RTP}. The present dielectric-loaded
device does not target the same aperture class; instead, it uses micron-scale optical confinement
and electrode spacing to reduce the drive voltage required for phase modulation while preserving the
bulk-crystal fabrication philosophy of RTP.
For the 10~mm bulk-cell example, the half-wave-voltage-length product is
$3.6~\mathrm{kV}\times1~\mathrm{cm}=3.6~\mathrm{kV\,cm}$, or $3600~\mathrm{V\,cm}$.

\begin{table*}[t]
\caption{Comparison between representative bulk RTP Pockels-cell operation and the proposed
dielectric-loaded RTP guided-wave phase modulator.}
\label{tab:bulk_comparison}
\centering
\small
\resizebox{\textwidth}{!}{%
\begin{tabular}{@{}p{0.25\linewidth}p{0.32\linewidth}p{0.34\linewidth}@{}}
\toprule
Metric & Bulk RTP Pockels cell & Proposed dielectric-loaded RTP waveguide \\
\midrule
Representative source & Raicol commercial RTP EO cell \cite{RaicolRtpEoCellWeb,RaicolRtpEoCellDatasheet}; RTP EO/Q-switch literature \cite{Albrecht2006RTP,Lebiush2000RTPQSwitch} & This work \\
Wavelength & 1064 nm & 1064 nm \\
Optical aperture / mode size & $1.5 \times 1.5$ to $15 \times 15$ mm$^2$ clear aperture & $A_\mathrm{eff}\approx1.11~\mu$m$^2$; $d_\mathrm{eff}\approx1.19~\mu$m \\
Interaction length & 10 mm example cell size; up to 50 mm available & Straight phase shifter; 5--10 mm representative length \\
Half-wave voltage & 3.6 kV for $9 \times 9 \times 10$ mm EO cell & $\approx9.7$ V for 5 mm; $\approx4.85$ V for 10 mm \\
$V_\pi L$ & 3.6 kV cm, or 3600 V cm, for 10 mm example & $\approx4.85$ V cm, single arm \\
Optical loss / transmission & $>98.5$\% transmission & Electrode-induced loss $\approx0.286$ dB/cm; estimated total propagation-loss budget $\approx1.3$ dB/cm, giving $\sim86$\% transmission for 5 mm and $\sim74$\% for 10 mm before coupling losses \\
Extinction ratio & $>30$ dB datasheet, $>35$ dB webpage & Phase-shifter metric only; extinction not evaluated \\
Drive-speed / ringing & $>1$ MHz, minimal ringing claimed & RF line designed near $50~\Omega$; estimated RC bandwidth $\approx19$ GHz for 1 mm \\
High-power suitability & High-power bulk device; commercial damage-threshold claim & $\Gamma_\mathrm{RTP}\approx48.6$\%; interface proxy tracked in Fig.~\ref{fig:design_tradeoffs} \\
Fabrication approach & Bulk crystal polishing and electrodes & $1.2~\mu$m-wide, 400~nm-thick SiN loading strip on bulk RTP; GSG electrodes with $g=5.25~\mu$m and $w_\mathrm{sig}=7.75~\mu$m \\
\bottomrule
\end{tabular}
}
\end{table*}

\section{Fabrication Tolerance and High-Power Considerations}
\label{sec:tolerance}

\subsection{Tolerance sweeps}

Fabrication sensitivity was evaluated by one-parameter perturbations around the selected geometry:
$w_\mathrm{SiN}=1.2~\mu\mathrm{m}$, $t_\mathrm{SiN}=400~\mathrm{nm}$,
$g=5.25~\mu\mathrm{m}$, and $w_\mathrm{sig}=7.75~\mu\mathrm{m}$. The tested ranges were
$w_\mathrm{SiN}=1.0$--$1.4~\mu\mathrm{m}$, $t_\mathrm{SiN}=350$--$450~\mathrm{nm}$,
$g=4.25$--$6.25~\mu\mathrm{m}$, and lateral electrode offsets of
$\pm0.5~\mu\mathrm{m}$. For each perturbation, the optical mode properties, active-axis fraction,
$\partial n_\mathrm{eff}/\partial V$, $V_\pi L$, $C'$, $Z_0$, and electrode-induced optical loss
were tracked. The EO/RF response is most sensitive to SiN thickness and electrode gap. Increasing
SiN thickness raises $V_\pi L$, while reducing electrode gap lowers $V_\pi L$ at the cost of
increased capacitance and electrode-induced optical loss. Electrode-gap and lateral-alignment
errors at the few-percent level produce only few-percent changes in the main EO/RF metrics, and
lateral electrode offsets of $\pm0.25~\mu\mathrm{m}$ have a modest effect on $V_\pi L$, $C'$, and
$Z_0$. Thus the selected design is not a fragile isolated point: it remains stable against
realistic electrode imperfections, while the SiN dimensions are the fabrication parameters that
most strongly control the optical confinement and should receive the tighter process control.

A practical fabrication consideration is the thermal budget associated with depositing SiN on RTP.
Stoichiometric LPCVD SiN can require temperatures approaching 700--800$^\circ$C, which may be
unnecessarily aggressive for a bulk RTP substrate because thermal-expansion mismatch and
temperature-dependent dielectric or ferroelectric properties can introduce film stress, adhesion
risk, or material changes~\cite{Perumal2020RTPDielectric}. For this reason, the intended process
window is better matched to lower-temperature PECVD or sputtered SiN deposition, followed by a
conservative post-deposition anneal only if needed. The present study therefore treats the SiN
layer as an optical loading element and does not assume a high-temperature LPCVD process;
experimental validation should include film-stress and adhesion tests on polished RTP witness
samples.

\subsection{High-power proxy metrics}

Direct optical-damage prediction is not attempted unless reliable material data are available for
the full device stack. Instead, the high-power behavior is interpreted through proxy metrics,
consistent with the role of bulk RTP in high-power EO-cell and Q-switch applications
\cite{Albrecht2006RTP,Lebiush2000RTPQSwitch}:
effective mode area, fraction of optical power in bulk RTP, fraction in SiN, fraction within
50~nm of the RTP/SiN interface, fraction near metal, and normalized peak intensity at material
interfaces. These quantities indicate whether the optical field remains concentrated primarily in
the bulk RTP region or is shifted into locations that are more sensitive to interface loss, heating,
or damage.

This proxy-metric approach supports the high-power motivation of the device without
overstating a quantitative damage threshold.

\section{Conclusion}

We have proposed a dielectric-loaded guided-wave EO modulator that uses a SiN strip on unetched
bulk RTP to create a near-surface guided mode while preserving the material advantages of RTP. The
study combines optical confinement, electrostatic field analysis, tensor-aware electro-optic
perturbation, comparison with bulk RTP Pockels-cell operation, and RF/device-level analysis to
assess whether a straight guided-wave RTP phase modulator can provide useful modulation efficiency
together with low capacitance and a bulk-compatible fabrication route. This platform is
particularly relevant to high-power, pulsed, and intracavity photonics, where electrical loading,
acoustic cleanliness, and optical robustness matter alongside EO efficiency.

\section*{Funding}

\subsection*{Acknowledgment}
Not applicable.

\subsection*{Disclosures}
The authors declare no conflicts of interest.

\subsection*{Data availability}
Data underlying the results presented in this paper are not publicly available at this time but may
be obtained from the authors upon reasonable request.

\bibliographystyle{unsrt}
\bibliography{refs}

\end{document}